\documentclass[twocolumn,showpacs,preprintnumbers,amsmath,amssymb,prb]{revtex4}
\usepackage{graphicx}
\usepackage{dcolumn}
\usepackage{bm}
\usepackage{braket}
\begin{document}

\title{Roles of Hund's rule coupling in excitonic density-wave states}

\author{Tatsuya Kaneko and Yukinori Ohta}
\affiliation{Department of Physics, Chiba University, Chiba 263-8522, Japan}


\date{Received 17 July 2014}

\begin{abstract}
Excitonic density-wave states realized by the quantum condensation of 
electron-hole pairs (or excitons) are studied in the two-band Hubbard 
model with Hund's rule coupling and the pair hopping term.  
Using the variational cluster approximation, we calculate the grand 
potential of the system and demonstrate that Hund's rule coupling 
always stabilizes the excitonic spin-density-wave state and destabilizes 
the excitonic charge-density-wave state and that the pair hopping term 
enhances these effects.  
The characteristics of these excitonic density-wave states are discussed 
using the calculated single-particle spectral function, density of states, 
condensation amplitude, and pair coherence length.  Implications of our 
results in the materials' aspects are also discussed.  
\end{abstract}


\maketitle

\section{Introduction}

The excitonic phases, often referred to as the excitonic insulator 
states or excitonic density-wave states, are described by the quantum 
condensation of excitons, which were predicted to occur in a small 
band-gap semiconductor or a small band-overlap semimetal.
\cite{jerome,halperin1,halperin2}
The exciton condensation in semimetallic systems can be described 
in analogy with the BCS theory of superconductors, and that in 
semiconducting systems can be discussed in terms of the Bose-Einstein 
condensation (BEC) of preformed excitons.\cite{bronold}  
The crossover phenomena between the BCS and the BEC states are then 
expected to produce rich physics in the field of quantum many-body 
systems. 
The excitonic phases are characterized by an order parameter 
$\langle c^\dag_{\bm{k+Q}\sigma}f_{\bm{k}\sigma'}\rangle$, where 
$c^\dag_{\bm{k}\sigma}$ and $f^\dag_{\bm{k}\sigma}$ are the creation 
operators of an electron in the conduction and valence bands, 
respectively.  If the valence-band top and conduction-band bottom 
are separated by the wave vector $\bm{Q}$, the system shows the 
density wave with modulation $\bm{Q}$.\cite{halperin1,halperin2}  
\\\indent
A number of candidate materials for the excitonic phases have been 
discovered.  It was claimed that Tm(Se,Te) shows a transition into 
the excitonic insulator state by applying pressure.\cite{bucher}  
The weak ferromagnetism of Ca$_{1-x}$La$_x$B$_6$ was interpreted 
in terms of a doped spin-triplet excitonic insulator.
\cite{young,zhitomirsky,balents}  Recently, the phase transition of 
a layered chalcogenide Ta$_2$NiSe$_5$ has been attributed to a 
realization of the spin-singlet excitonic insulator and has attracted 
much experimental and theoretical attention.
\cite{wakisaka1,wakisaka2,kaneko1,seki1} 
The charge-density wave (CDW) of $1T$-TiSe$_2$ has also been claimed 
to be of the excitonic origin.\cite{cercellier,monney1,monney2,zenker1} 
The spin-density wave (SDW) state of iron pnictide superconductors 
has sometimes been argued to be of the excitonic origin as well.
\cite{brydon1,brydon2,zocher} 
It was proposed that the condensation of spin-triplet excitons 
can occur in the proximity of a spin-state transition;\cite{kunes1} 
Pr$_{0.5}$Ca$_{0.5}$CoO$_3$ is an example.\cite{kunes2} 
\\\indent
In this paper, motivated by the above development in the field, 
we study the stability of the excitonic density-wave states in 
the two-band Hubbard model where Hund's rule coupling $J$,  
the pair hopping term $J'$, as well as the interorbital Coulomb 
repulsion $U'$ are taken into account in addition to the standard 
intraorbital Hubbard repulsion $U$.  
It is known that the interorbital repulsion $U'$ induces the 
excitonic instability in the system,\cite{zocher,kaneko2} but the 
condensations of the spin-singlet and spin-triplet excitons 
are exactly degenerate unless Hund's rule coupling, the pair hopping 
term, or electron-phonon coupling are taken into account.  Thus, 
we here study the roles of Hund's rule coupling and the pair hopping 
term played in the excitonic density wave of which 
not much is known so far.  
\\\indent
We first rewrite the interorbital interaction terms of the 
Hamiltonian in terms of the creation and annihilation operators 
of the spin-singlet and spin-triplet excitons.  We then show 
that the interorbital repulsion $U'$ actually leads to the 
exciton formation in both the spin-singlet and the spin-triplet channels 
and that Hund's rule coupling always lowers the energy of 
the spin-triplet exciton and raises the energy of the spin-singlet 
exciton.  
The variational cluster approximation (VCA)\cite{potthoff1,dahnken} 
is then used to study the two-band Hubbard model in detail, and 
we show that Hund's rule coupling and the pair hopping term always 
stabilize the excitonic SDW state and destabilize the excitonic 
CDW state.  
The characteristics of these excitonic density-wave states 
will moreover be examined using a variety of calculated physical 
quantities, including the single-particle spectral function, density 
of states (DOS), condensation amplitude, and pair coherence length.  
Consequences of the present results on the excitonic density-wave 
states of a variety of materials will also be discussed.  
\\\indent
This paper is organized as follows:  In Sec.~II, the model and 
method of calculations will be given.  In Sec.~III, the results 
of calculations for various physical quantities will be 
presented.  Summary and discussion will be given in Sec.~IV.

\section{Model and Method}

\subsection{The two-band Hubbard model} 

We consider the two-band Hubbard model defined by the Hamiltonian,
\begin{align}
\mathcal{H}=&-t\sum_{\langle i,j\rangle}\sum_{\sigma}\sum_{\alpha}\alpha^{\dag}_{i\sigma}\alpha_{j\sigma} 
-D\sum_{i}(n_{if}-n_{ic}) \notag \\ 
&+U\sum_{i}\sum_{\alpha} n_{i\alpha\uparrow}n_{i\alpha\downarrow}+U'\sum_{i}n_{if}n_{ic} \notag \\ 
&-2J\sum_{i}\left( \bm{S}_{if}\cdot\bm{S}_{ic}+\frac{1}{4}n_{if}n_{ic} \right) \notag \\
&-J'\sum_{i}\left( f^{\dag}_{i\uparrow}f^{\dag}_{i\downarrow}c_{i\uparrow}c_{i\downarrow}
+c^{\dag}_{i\uparrow}c^{\dag}_{i\downarrow}f_{i\uparrow}f_{i\downarrow} \right),
\label{ham}
\end{align} 
where $\alpha^{\dag}_{i\sigma}$ $(=f^{\dag}_{i\sigma},c^{\dag}_{i\sigma})$ 
denotes the creation operator of an electron with spin 
$\sigma$ $(=\uparrow,\downarrow)$ 
on the $\alpha$ $(=f,c)$ orbital at site $i$ and 
$n_{i\alpha}=n_{i\alpha\uparrow}+n_{i\alpha\downarrow}
=\alpha^\dag_{i\uparrow}\alpha_{i\uparrow}+\alpha^\dag_{i\downarrow}\alpha_{i\downarrow}$. 
$t$ is the hopping integral between the same orbitals on the neighboring sites, 
and $D$ is the level splitting between the two orbitals.  
$U$ and $U'$ are the intra- and interorbital Coulomb repulsions, respectively, between electrons and 
$J$ and $J'$ are the strengths of Hund's rule coupling and the pair hopping term, respectively.  
Note that $J=J'$ in the standard two-orbital Hubbard model, but when necessary 
we retain only Hund's rule coupling by setting $J'=0$ to examine the role 
of the pair hopping term.  
Throughout the paper, we assume the filling of two electrons per site (or half-filling), i.e., 
$\langle n_{f}\rangle+\langle n_{c}\rangle=2$, where 
$\langle n_{\alpha}\rangle=\sum_{i,\sigma}\langle n_{i\alpha\sigma}\rangle/N$.

The Hamiltonian Eq.~(\ref{ham}) in the spinless case with $U=J=J'=0$ is 
equivalent to the extended Falicov-Kimball model with dispersive $c$ 
and $f$ electrons of which the excitonic insulator state has been studied 
much in detail.\cite{batista,seki2,zenker2,kaneko3,ejima}
The excitonic phases in the two-band Hubbard model without Hund's rule 
coupling and the pair hopping term ($J=J'=0$) have also been studied 
recently\cite{zocher,kaneko2} where it was shown that the model exhibits 
three ground-state phases: 
(i) the band insulator (at $U',D\gg U,J$), where 
$\langle n_{f}\rangle =2$ and $\langle n_{c}\rangle=0$, 
(ii) the antiferromagnetic Mott insulator (at $U,J\gg U'D$), where 
$\langle n_{f}\rangle=\langle n_{c}\rangle = 1$, and 
(iii) the excitonic density-wave state between the above two phases, 
where $2>\langle n_{f}\rangle>1>\langle n_{c}\rangle>0$.  
However, although many studies have been performed on the multiband 
Hubbard models, recently in relation to iron pnictide superconductors,
\cite{dagotto,brydon,daghofer,luo} the excitonic density-wave states 
in the two-band Hubbard model with Hund's rule coupling have not 
been studied in detail; only a recent dynamical mean-field theory 
(DMFT) calculation\cite{kunes1,kunes2} is noticed.  

To see the stability of the spin-singlet and spin-triplet excitons, 
let us introduce the creation operators of the spin-singlet and spin-triplet 
excitons, respectively, which are defined as 
\begin{align}
{A^0_{i}}^{\dag} = \frac{1}{\sqrt{2}}\sum_{\sigma}c^{\dag}_{i\sigma}f_{i\sigma}, \;\;\;
{\bm{A}_{i}}^{\dag} = \frac{1}{\sqrt{2}}\sum_{\sigma\sigma'}c^{\dag}_{i\sigma}\bm{\sigma}_{\sigma\sigma'}f_{i\sigma'},
\end{align}
where $\bm{\sigma}$ is the vector of the Pauli matrices.  
Using the spin-singlet and spin-triplet exciton operators thus defined, 
the interorbital Coulomb repulsion term can be divided exactly 
into the spin-singlet and spin-triplet terms as 
\begin{align}
U' n_{if}n_{ic} = -U'{A^0_{i}}^{\dag} A^0_{i} -U'\bm{A}^{\dag}_{i}\cdot\bm{A}_{i} + 2U' n_{ic}.
\end{align}
Therefore, the formation of excitons lowers the energy of the system 
in both the spin-singlet and the spin-triplet channels by the same amount.  
Hund's rule coupling and the pair hopping terms can also be rewritten 
exactly as
\begin{align}
-J \bm{S}_{if}\cdot\bm{S}_{ic}&=\frac{3J}{4}{A^0_{i}}^{\dag}A^0_{i}-\frac{J}{4}\bm{A}^{\dag}_{i}\cdot\bm{A}_{i}, \\
-J'c^{\dag}_{i\uparrow}c^{\dag}_{i\downarrow}f_{i\uparrow}f_{i\downarrow}&=\frac{J'}{4}{A^0_{i}}^{\dag}{A^0_{i}}^{\dag} 
-\frac{J'}{4} \bm{A}^{\dag}_{i}\cdot\bm{A}^{\dag}_{i}. \label{pair} 
\end{align}
Therefore, due to the Hund's rule coupling term, the formation of 
the spin-triplet (spin-singlet) excitons always lowers (raises) 
the energy of the system, thus lifting the degeneracy that occurs 
at $J=J'=0$.  The pair hopping term can also be divided into the 
spin-singlet and spin-triplet terms as in Eq.~(\ref{pair}), which are 
of the off-diagonal form.  

\subsection{Variational cluster approximation} 

We use the VCA,\cite{potthoff1,dahnken} which is a quantum cluster method 
based on the self-energy functional theory,\cite{potthoff2} and 
solve the quantum many-body problem defined in Eq.~(\ref{ham}).  
Note that, unlike in DMFT, we can taken into account the effects of 
short-range spatial electron correlations precisely in this approach.  
The VCA introduces the disconnected finite-size clusters that are solved 
exactly to obtain the exact self-energy of the clusters $\hat{\Sigma}'$ 
with which a superlattice is formed as a reference system.  
The matrices are indicated by a $\hat{~}$ hereafter.  
By restricting the trial self-energy to $\hat{\Sigma}'$, we obtain 
an approximate grand potential of the original system, 
\begin{align}
\Omega=\Omega'-\frac{1}{N}\oint_{C}\frac{{\rm d}z}{2\pi i}\sum_{\bm{K},\sigma}\ln\det
\big[\hat{I}-\hat{V}_{\sigma}(\bm{K})\hat{G}'_{\sigma}(z)\big], \label{gp}
\end{align}
where $\Omega'$ is the grand potential of the reference system, 
$\hat{I}$ is the unit matrix, 
$\hat{V}$ is the hopping matrix between adjacent clusters, and 
$\hat{G}'$ is the exact Green's function of the reference system.  
The $\bm{K}$ summation is performed in the reduced Brillouin zone of 
the superlattice, and the contour $C$ of the frequency integral encloses 
the negative real axis.  
Details of the VCA can be found in Refs.~[\onlinecite{potthoff0,senechal0}].
 
To study the symmetry-breaking phases in the VCA, we introduce the Weiss 
fields as variational parameters.  The variational Hamiltonian for the 
excitonic CDW and SDW states are then defined as 
\begin{align}
&\mathcal{H}'_{\mathrm{CDW}}=\mathcal{H}+\Delta'_0 \sum_{i,\sigma} 
e^{i\bm{Q}\cdot\bm{r}_i}\big(c^{\dag}_{i\sigma}f_{i\sigma}+\mathrm{H.c.}\big), \label{vc} \\
&\mathcal{H}'_{\mathrm{SDW}}=\mathcal{H}+\Delta'_z\sum_{i,\sigma}\sigma 
e^{i\bm{Q}\cdot\bm{r}_i}\big(c^{\dag}_{i\sigma}f_{i\sigma}+\mathrm{H.c.}\big), \label{vs}
\end{align} 
respectively, where $\Delta'_0$ is the Weiss field for condensation 
of the spin-singlet excitons and $\Delta'_z$ is the $z$ component of the 
Weiss field for condensation of the spin-triplet excitons.  
The variational parameters $\Delta'_0$ and $\Delta'_z$ are optimized 
on the basis of the variational principle, i.e., 
$\partial\Omega/\partial\Delta'_0=0$ for the excitonic CDW state and 
$\partial\Omega/\partial\Delta'_z=0$ for the excitonic SDW state.  
The solutions with $\Delta'_0 \ne 0$ and $\Delta'_z\ne 0$ correspond to 
the excitonic CDW and SDW states, respectively.  

We solve the eigenvalue problem $\mathcal{H}'|\psi_0\rangle = E_0|\psi_0\rangle$ 
of a finite-size ($L_c$ sites) cluster to obtain the ground state, and 
we calculate the trial Green's function by the Lanczos exact-diagonalization 
method.  
Using the basis $\bm{\Psi}^{\dag}_{i} = (f^{\dag}_{i\sigma},c^{\dag}_{i\sigma})$, 
the Green's-function matrix $\hat{G}'_\sigma$ in Eq.~(\ref{gp}) may 
be written as 
\begin{align}
\hat{G}'_{\sigma}(\omega )
=\left( \begin{array}{cc}
\hat{G}'^{ff}_{\sigma}(\omega ) &\hat{G}'^{fc}_{\sigma}(\omega ) \\
\hat{G}'^{cf}_{\sigma}(\omega ) &\hat{G}'^{cc}_{\sigma}(\omega )  \\
\end{array} \right), 
\end{align}
where $\hat{G}'^{\alpha\beta}_{\sigma}$ is an $L_c \times L_c$ matrix 
and each matrix element is defined as 
\begin{align}
G'^{\alpha\beta}_{ij,\sigma}(\omega) &
= \langle \psi_0 | \alpha_{i\sigma}\frac{1}{\omega-\mathcal{H}'+E_0}\beta^{\dag}_{j\sigma} |\psi_0\rangle  \nonumber\\
&+ \langle \psi_0 | \beta^{\dag}_{j\sigma} \frac{1}{\omega+\mathcal{H}'-E_0}\alpha_{i\sigma} |\psi_0\rangle.
\end{align} 
The matrix $\hat{V}$ in Eq.~(\ref{gp}) is given as 
\begin{align}
\hat{V}_{\sigma}(\bm{K})=\left(
\begin{array}{cc}
\hat{T}(\bm{K})&-\Delta'_{\sigma}\hat{I}\\
-\Delta'_{\sigma}\hat{I}& \hat{T}(\bm{K})\\
\end{array} 
\right),
\end{align}
where $\hat{T}(\bm{K})$ is the intercluster hopping matrix with the matrix elements 
$T_{ij}(\bm{K} )=-t \sum_{\bm{X},x}e^{i\bm{K}\cdot\bm{X}}\delta _{i+x,j}\delta_{\bm{R}+\bm{X},\bm{R}'}$.  
Here, $x$ denotes the neighboring site of the $i$ th site, and $\bm{X}$ denotes 
the neighboring cluster of the $\bm{R}$ th cluster. 
$\Delta'_{\sigma}=\Delta'_0$ for the excitonic CDW state, and 
$\Delta'_{\sigma}=\sigma\Delta'_z$ for the excitonic SDW state.  

In our VCA calculation, we assume the two-dimensional square lattice 
and use an $L_c=2\times2=4$ site (eight-orbital) cluster as the reference system.  
We set $D/t=3.2$ so that the noninteracting tight-binding band structure 
is a  semimetal with a small band overlap.  
The band structure has an electron pocket at $\bm{k}=(0,0)$ and 
a hole pocket at $\bm{k}=(\pi,\pi)$ of the Brillouin zone. 
Hence, the modulation vector of the density waves is given by 
$\bm{Q}=(\pi,\pi)$.  
Throughout the paper, we assume the relation $U'=(U+J)/2$ between 
the interaction parameters so that the Hartree shift can be 
suppressed.  The standard choice $U'=U-2J$ (with $J'=J$) valid in the 
atomic limit\cite{griffith,brandow,oles} has also been used to check 
that the essential features obtained in the present paper do not 
change (see the Appendix).  We moreover assume the value $U/t=8$ at 
which the excitonic density-wave state is stabilized between the 
band-insulator and the Mott-insulator states.\cite{kaneko2}  

\begin{figure}[!t]
\begin{center}
\includegraphics[width=\columnwidth]{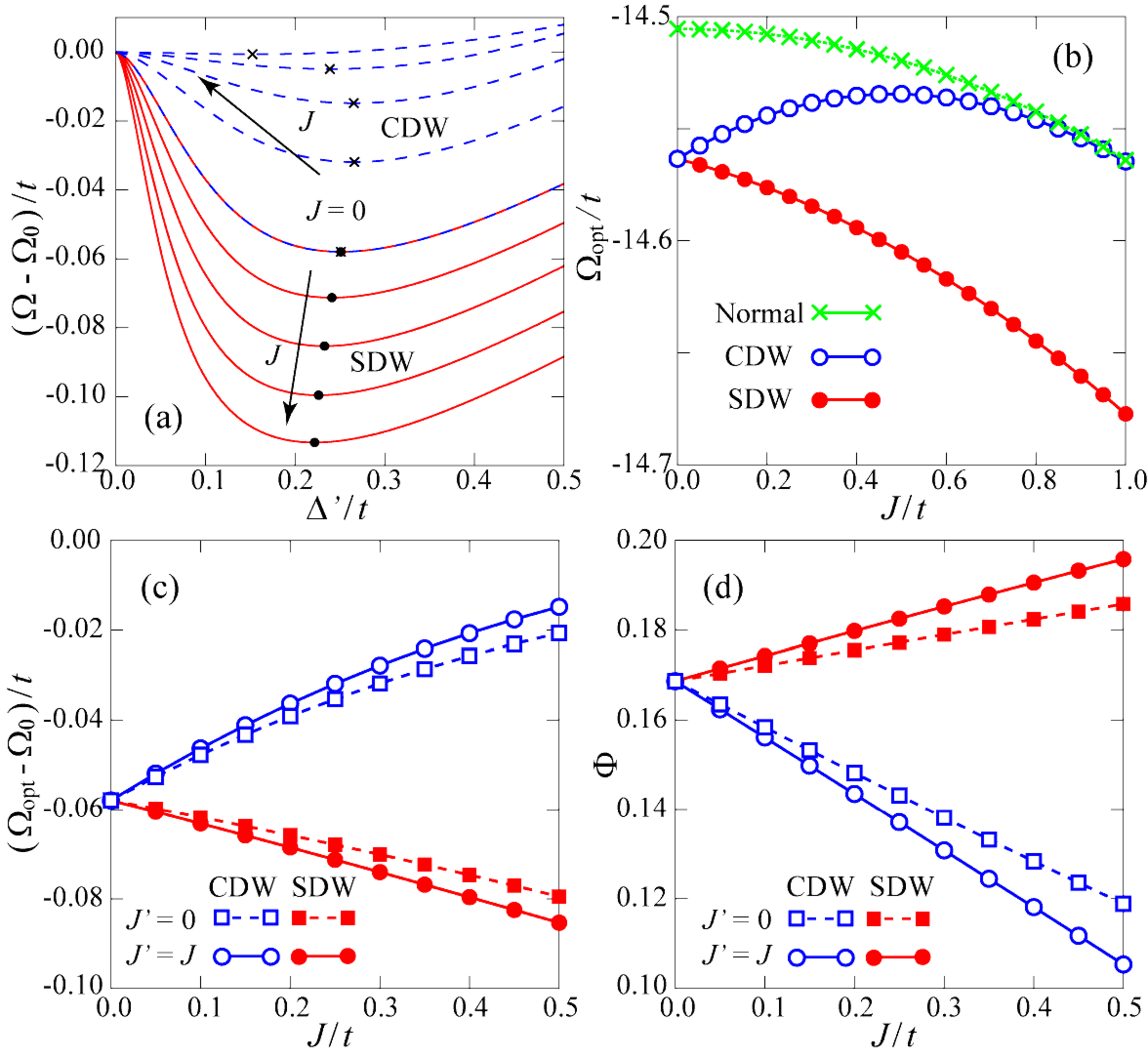}
\caption{(Color online) (a) Calculated grand potentials for the excitonic 
CDW and SDW states as a function of the variational parameter $\Delta'$ 
($=\Delta'_0,\Delta'_z$) at $J/t=J'/t=0$, $0.25$, $0.5$, $0.75$, and $1$.  
$\Omega_0$ is the grand potential in the normal (semimetallic) state.  
The crosses and dots indicate the stationary points of the excitonic CDW 
and SDW states, respectively.  
(b) $J$ ($=J'$) dependence of the grand potential at the stationary point 
for the normal (or semimetallic), excitonic CDW, and SDW states.  
(c) Optimized values of the grand potentials in the presence ($J'=J$) and 
absence ($J'=0$) of the pair hopping term.  
(d) $J$ dependence of the order parameters of the excitonic CDW and SDW 
states in the presence ($J'=J$) and absence ($J'=0$) of the pair hopping 
term.  
}\label{fig1}
\end{center}
\end{figure}

\section{Results of calculation}

\subsection{Stability of the excitonic density waves}

First, let us examine the stability of the excitonic CDW and SDW 
states using the grand potential.  In Fig.~\ref{fig1}(a), we show 
the calculated grand potentials of the excitonic CDW and SDW states 
as a function of the variational parameter $\Delta'$.  
We find that the grand potential has the stationary points at 
$\Delta'=0$ and $\Delta'\ne 0$, and the latter is lower in energy, 
indicating that the excitonic density-wave states are 
thermodynamically stable.  At $J=J'=0$, the grand potentials of 
the excitonic CDW and SDW states are exactly degenerate 
[see Fig.~\ref{fig1}(a)], but Hund's rule coupling $J$ and 
pair hopping term $J'$ lift this degeneracy.  
The optimized values of the grand potential as a function of $J$ 
(=$J'$) are shown in Fig.~\ref{fig1}(b) where we find that, with 
increasing $J$ (and $J'$), the energy of the excitonic SDW state 
decreases, but the energy of the excitonic CDW state increases and 
approaches the energy of the normal semimetallic state.  
Therefore, the excitonic SDW (CDW) state is stabilized (destabilized) 
by $J$ and $J'$.  
In Fig.~\ref{fig1}(c), we show the optimized values of the grand 
potentials in the presence ($J'=J$) and absence ($J'=0$) of the 
pair hopping term where we find that the stability of the 
excitonic SDW (CDW) state is enhanced (suppressed) by pair 
hopping term $J'$.   

We also calculate the order parameters of the excitonic CDW and 
SDW states.  Here, we introduce the quantities $\Phi_0$ and $\Phi_z$ 
for the excitonic CDW and SDW order parameters, respectively, 
which are defined as 
\begin{align}
&\Phi_0 = \frac{1}{2N}\sum_{\bm{k}}\sum_{\sigma} 
\langle c^{\dag}_{\bm{k}+\bm{Q}\sigma}f_{\bm{k}\sigma} \rangle, \\
&\Phi_z = \frac{1}{2N}\sum_{\bm{k}}\sum_{\sigma} 
\sigma \langle c^{\dag}_{\bm{k}+\bm{Q}\sigma}f_{\bm{k}\sigma} \rangle. 
\end{align}
The calculated results for $\Phi_0$ and $\Phi_z$ are shown in 
Fig.~\ref{fig1}(d) in the presence ($J'=J$) and absence ($J'=0$) of 
the pair hopping term.  We find that $\Phi_z$ is enhanced with $J$ 
(and $J'$) and $\Phi_0$ is suppressed with $J$ (and $J'$), which 
are in accordance with the stability of the excitonic CDW and SDW 
states evaluated from the behaviors of the calculated grand 
potentials.  
Thus, we may state that Hund's rule coupling stabilizes the 
excitonic SDW state and destabilizes the excitonic CDW state.  
As seen in Figs.~\ref{fig1}(c) and \ref{fig1}(d), we may moreover state that 
the pair hopping term enhances the stability of the excitonic SDW 
state and suppresses the stability of the excitonic CDW state.

\begin{figure}[htb]
\begin{center}
\includegraphics[width=\columnwidth]{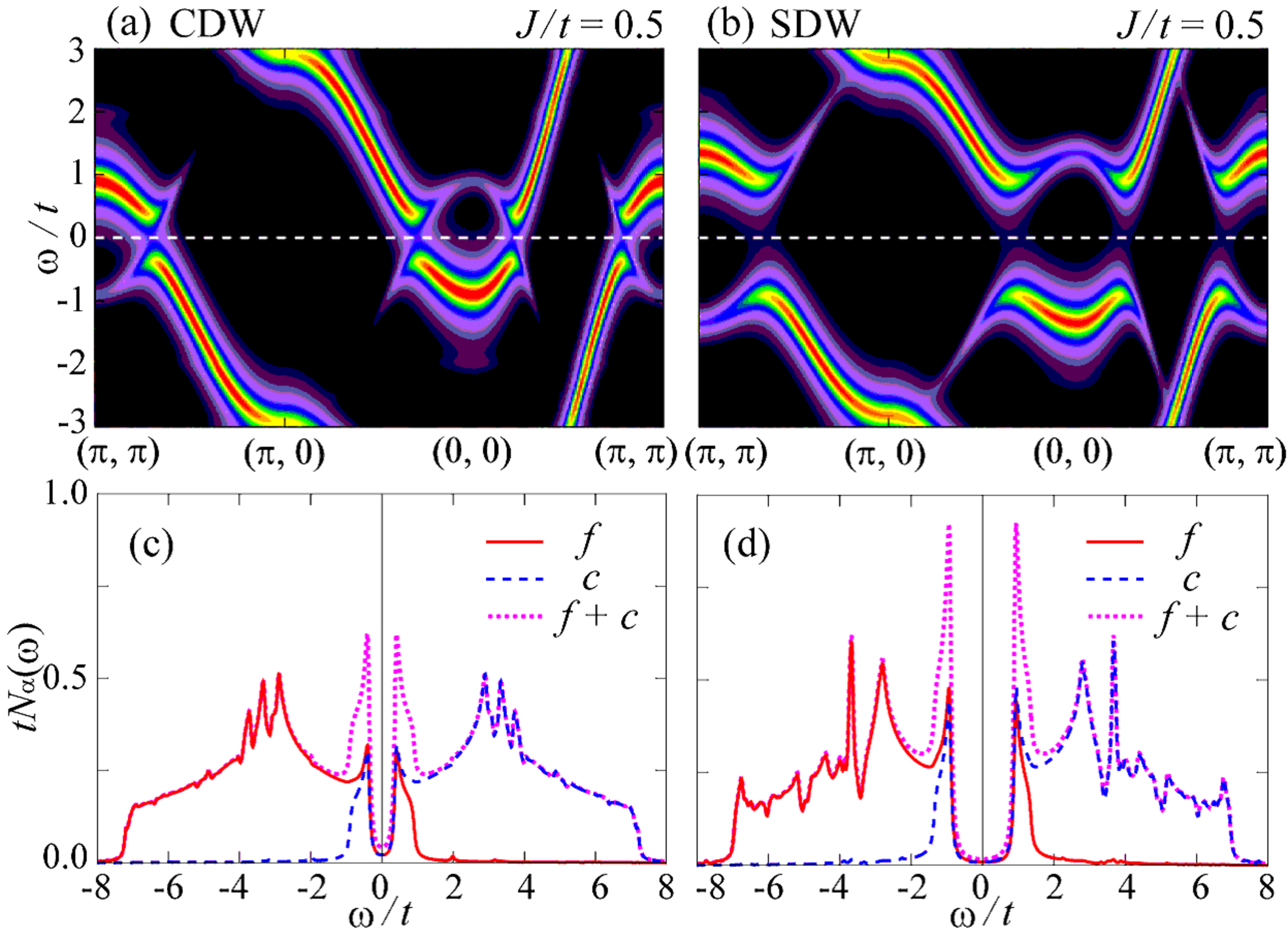}
\caption{(Color online) Single-particle spectral function $A(\bm{k},\omega)$ 
and DOS $N_{\alpha}(\omega)$ calculated by CPT at $J/t=J'/t=0.5$.  We show the 
results for the excitonic CDW state (metastable) in (a) and (c) and for 
the excitonic SDW state (stable) in (b) and (d).  In (c) and (d), the solid, 
dashed, and dotted lines indicate the $f$ orbital, $c$ orbital, and total 
DOSs, respectively.  The artificial Lorentzian broadening of $\eta/t=0.15$ 
is used for $A(\bm{k},\omega)$ and $\eta/t=0.05$ is used for $N_{\alpha}(\omega)$.  
The Fermi level is located at $\omega=0$.  
}\label{fig2}
\end{center}
\end{figure}

\subsection{Single-particle spectral function}

Next, let us calculate the Green's function at the optimized values 
of the variational parameters using the cluster perturbation 
theory (CPT).\cite{senechal}  The Green's function is defined as
\begin{equation}
\hat{\mathcal{G}}_{\sigma}(\bm{k},\bm{k}',\omega)
=\frac{1}{L_c}\sum^{L_c}_{i,j=1}
\hat{G}^{\mathrm{CPT}}_{ij,\sigma}({\bm{k}},\omega)
e^{-i{\bm{k}\cdot\bm{r}_i}+i\bm{k}'\cdot\bm{r}_j}, \label{CPT}
\end{equation}
where $\hat{G}^{\mathrm{CPT}}_{\sigma}(\bm{k},\omega)
=\big[\hat{G}'^{-1}_{\sigma}(\omega)-\hat{V}_{\sigma}(\bm{k})\big]^{-1}$.  
Using this Green's function, the single-particle spectral function 
is defined as 
\begin{align}
A(\bm{k},\omega)=-\frac{1}{\pi}\;\sum_{\alpha,\sigma}
\mathrm{Im}\; \mathcal{G}^{\alpha\alpha}_{\sigma}(\bm{k},\bm{k},\omega+i\eta), 
\label{SP}
\end{align}
where $\eta$ gives the artificial Lorentzian broadening to the spectrum.  
We also calculate the DOS for the $\alpha$ $(=f,c)$ 
orbital, which is defined as
\begin{align}
N_{\alpha}(\omega)=-\frac{1}{\pi N} \sum_{\bm{k}}\sum_{\sigma}
\mathrm{Im}\; \mathcal{G}^{\alpha\alpha}_{\sigma}(\bm{k},\omega+i\eta).
\end{align}

\begin{figure}[htb]
\begin{center}
\includegraphics[width=\columnwidth]{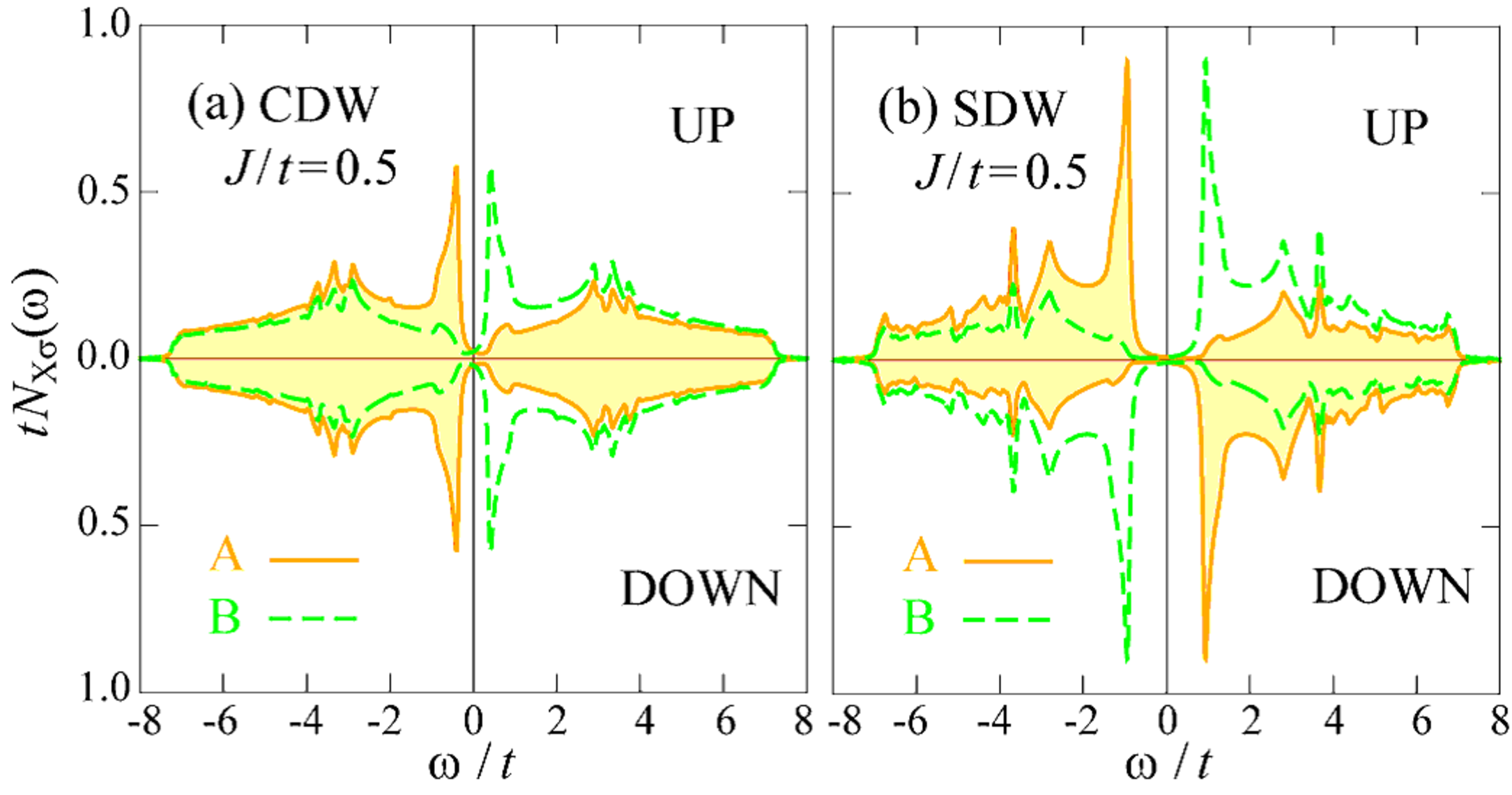}
\caption{(Color online) Calculated DOSs for the (a) excitonic CDW state and 
(b) excitonic SDW state at $J/t=J'/t=0.5$.  Solid and dashed lines indicate 
the DOSs of the A and B sublattices, respectively.  The Lorentzian broadening 
of $\eta/t=0.05$ is used.  The vertical line indicates the Fermi level.  
}\label{fig3}
\end{center}
\end{figure}

\begin{figure*}[!t]
\begin{center}
\includegraphics[width=16.0cm]{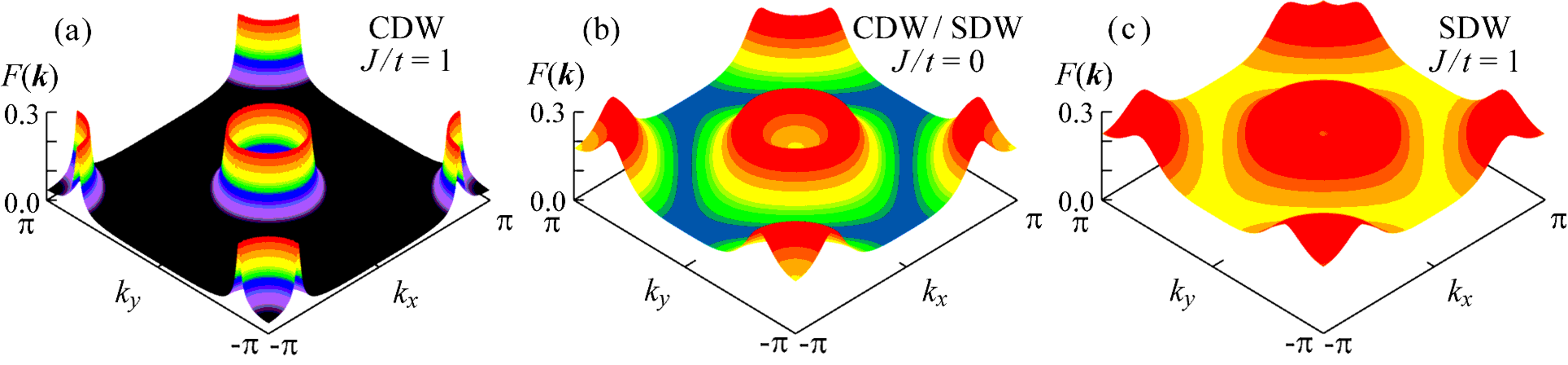}
\caption{(Color online) Condensation amplitude $F(\bm{k})$ [$=F_0(\bm{k})$ 
or $F_z(\bm{k})$] calculated by CPT.  We show the results for 
(a) the CDW state at $J/t=J'/t=1.0$ (metastable), 
(b) the CDW/SDW states at $J=J'=0$ (degenerate), and 
(c) the SDW state at $J/t=J'/t=1.0$ (stable).}
\label{fig4}
\end{center}
\end{figure*}

In Fig.~\ref{fig2}, we show the calculated single-particle spectral 
function $A(\bm{k},\omega)$ and DOS $N_{\alpha}(\omega)$; 
the results for the metastable CDW state [see Figs.~\ref{fig2}(a) and \ref{fig2}(c)] 
and stable SDW state [see Figs.~\ref{fig2}(b) and \ref{fig2}(d)] obtained at 
$J/t=J'/t=0.5$ are shown.  We find that, although a semimetallic state with 
a small band overlap is assumed as the noninteracting band structure, the 
valence band around $\bm{k}=(\pi, \pi)$ is hybridized with the conduction 
band around $\bm{k}=(0,0)$ due to the spontaneous $c$-$f$ hybridization 
(or exciton condensation), leading to the opening of the band gap 
at the Fermi level. 
At $J=J'=0$, the single-particle excitation gap $\Delta_g$ is estimated 
to be $\Delta_g/t=1.47$.  We find that, in agreement with the change 
in the order parameters, the single-particle gap in the excitonic CDW state, 
e.g., $\Delta_g/t=0.76$ at $J/t=J'/t=0.5$, is suppressed in comparison with 
the $J=J'=0$ case [see Figs.~\ref{fig2}(a) and \ref{fig2}(c)].  
We also find that the single-particle gap in the excitonic SDW state, e.g., 
$\Delta_g/t=1.81$ at $J/t=J'/t=0.5$, is enhanced in comparison with the 
$J=J'=0$ case [see Figs.~\ref{fig2}(b) and \ref{fig2}(d)].  
We moreover find in Figs.~\ref{fig2} (c) and \ref{fig2}(d) that the 
sharp coherence peak appears at the edges of the gap and that the 
coherence peak of the SDW state is sharper than that of the CDW state, 
indicating that the spontaneous $c$-$f$ hybridization in the excitonic SDW (CDW) 
state is enhanced (suppressed) by Hund's rule coupling and the pair 
hopping term.  
We note that no significant differences are found in the behaviors of 
$N_{\alpha}(\omega)$ discussed above, even if we switch off the pair 
hopping term, retaining only Hund's rule coupling.  

In order to see the character of the excitonic density-wave states, 
we calculate the DOS of the A and B sublattices.  The sublattice 
Green's function is given by
\begin{align}
\hat{\mathcal{G}}_{X\sigma}(\bm{k},\omega)
=\frac{2}{L_c}\sum_{i,j\in X}
\hat{G}^{\mathrm{CPT}}_{ij,\sigma}({\bm{k}},\omega)
e^{-i{\bm{k}\cdot( \bm{r}_i}-\bm{r}_j)}  \label{subG}
\end{align}
with $X=$ A or B.  Using this sublattice Green's function, the DOS 
of the A or B sublattices is defined as 
\begin{align}
N_{X\sigma}(\omega)=-\frac{1}{\pi N} \sum_{\bm{k}}\sum_{\alpha,\beta}
\mathrm{Im}\; \mathcal{G}^{\alpha\beta}_{X\sigma}(\bm{k},\omega+i\eta).
\end{align}

In Fig.~\ref{fig3}(a), we show the calculated DOS for the excitonic 
CDW state at $J/t=J'/t=0.5$.  We note that, below the Fermi level 
($\omega<0$), the up- and down-spin DOSs are the same and 
the DOS of the A sublattice is larger than that of the B sublattice: 
$N_{A\uparrow}(\omega)=N_{A\downarrow}(\omega) > N_{B\uparrow}(\omega)=N_{B\downarrow}(\omega)$. 
We also note that $N_{A\sigma}(\omega)\simeq N_{B\sigma}(\omega)$ 
far away from the Fermi level and that the coherence peak appears 
in the DOS of the A sublattice just below the Fermi level, where 
$N_{A\sigma}(\omega)> N_{B\sigma}(\omega)$.  Using the order 
parameter $\Phi_0$, the local number of electrons is given by 
$n_i = 1 + 2 \Phi_0 \cos \left( \bm{Q}\cdot\bm{r}_i \right)$.  
At $J/t=J'/t=0.5$, we have $\Phi_0=0.11$, and thus the local numbers 
of the electrons on each sublattice are given by 
$n_{A\uparrow}=n_{A\downarrow}=1+2\Phi_0=1.22$ and 
$n_{B\uparrow}=n_{B\downarrow}=1-2\Phi_0=0.78$.  
We therefore find that, due to the effect of Hund's rule coupling 
and the pair hopping term, $\Phi_0$ is suppressed and thus the excitonic 
CDW modulation in real space becomes rather weak.  

In Fig.~\ref{fig3}(b), we show the calculated DOS for the excitonic 
SDW state at $J/t=J'/t=0.5$.  We note that, below Fermi level, the up-spin 
DOS of the A (B) sublattice is equal to the down-spin DOS of the 
B (A) sublattice and that the up-spin DOS of the A (B) sublattice 
is larger (smaller) than the down-spin DOS of the A (B) sublattice: 
$N_{A\uparrow}(\omega)=N_{B\downarrow}(\omega) > N_{A\downarrow}(\omega)=N_{B\uparrow}(\omega)$. 
We also note that the DOS has a large gap and a sharp coherence 
peak appears at the edge of the DOS.  Using the order parameter 
$\Phi_z$, the local magnetization is given by 
$m_i = 2 \Phi_z \cos \left( \bm{Q}\cdot\bm{r}_i \right)$.  
At $J/t=J'/t=0.5$, we have $\Phi_z=0.20$, and thus the local numbers 
of electrons on each sublattice are given by 
$n_{A\uparrow}=n_{B\downarrow}=1+2\Phi_z=1.40$ and 
$n_{A\downarrow}=n_{B\uparrow}=1-2\Phi_z=0.60$.  
We therefore find that, due to the effect of Hund's rule coupling 
and the pair hopping term, $\Phi_z$ is enhanced, and thus the excitonic 
SDW modulation in real space becomes rather strong.

\subsection{Condensation amplitude and coherence length}

In order to see the character of the exciton condensation in momentum 
space, we calculate the condensation amplitude (or the anomalous 
momentum distribution function).  Using the off-diagonal (or anomalous) 
Green's function given in Eq.~(\ref{CPT}), the condensation amplitudes 
for the spin-singlet and spin-triplet excitons are defined as
\begin{align}
F_0({\bm k})&=\frac{1}{2}\sum_{\sigma}\oint_{C}
\frac{{\rm d}z}{2\pi i}
\mathcal{G}^{cf}_{\sigma}({\bm k},\bm{k}+\bm{Q},z), \\
F_z({\bm k})&=\frac{1}{2}\sum_\sigma\sigma\oint_{C}
\frac{{\rm d}z}{2\pi i}
\mathcal{G}^{cf}_{\sigma}({\bm k},\bm{k}+\bm{Q},z), 
\end{align}
respectively.  Note that we here use the term ``anomalous'' to indicate 
that the number of electrons on each of the $c$ and $f$ orbitals 
is not conserved due to the excitonic condensation, although the 
total number of electrons is conserved.  

We show the calculated results in Fig.~\ref{fig4} for the excitonic 
CDW and SDW states.  We find that, with increasing $J$ ($=J'$), the peak 
of $F_0(\bm{k})$ at the Fermi momentum $\bm{k}_{\rm F}$ becomes sharper in the 
CDW state [see Fig.~\ref{fig4}(a)] and that the peak of $F_z(\bm{k})$ 
at $\bm{k}_{\rm F}$ becomes broader in momentum space in the SDW state 
[see Fig.~\ref{fig4}(c)].  The sharp (broad) peak of $F(\bm{k})$ 
[$=F_0(\bm{k})$ or $F_z(\bm{k})$] in momentum space indicates that 
the spatial extension of the electron-hole pair becomes large (small) 
in real space.  We note that no significant differences are found in 
the behavior of $F(\bm{k})$, even if we set $J'=0$ retaining only  
Hund's rule coupling.  

\begin{figure}[!t]
\begin{center}
\includegraphics[width=0.9\columnwidth]{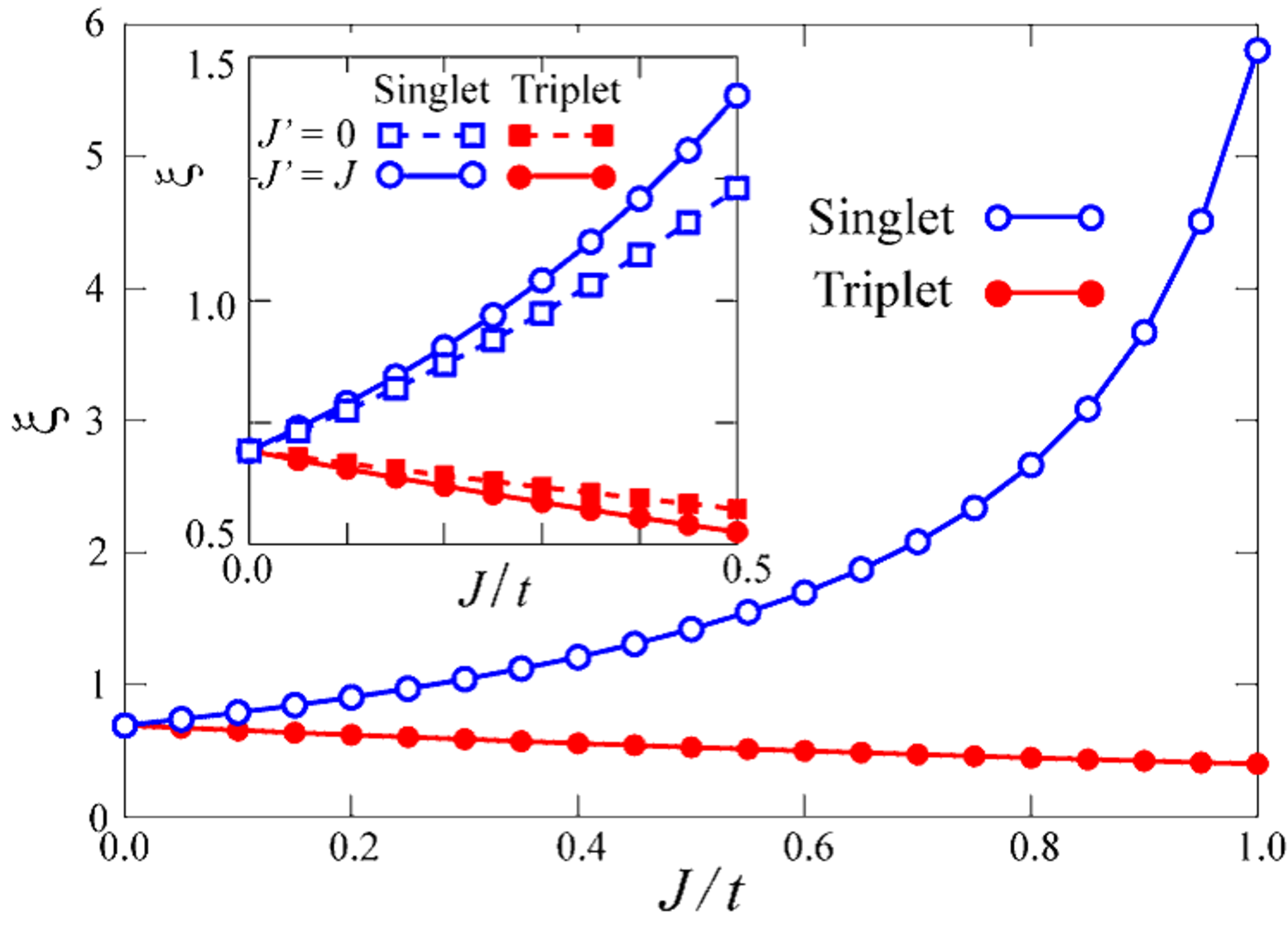}
\caption{(Color online) Calculated pair coherence length $\xi$ in 
units of the lattice constant.  $J$ ($=J'$) dependence of $\xi$ 
is shown for the spin-singlet (open circles) and spin-triplet 
(solid circles) exciton condensations.  The inset shows the 
results in the absence of the pair hopping term $J'=0$ (open and 
solid squares), which are compared with the results in the presence 
of the pair hopping term $J'=J$ (open and solid circles).  
}\label{fig5}
\end{center}
\end{figure}

Using $F(\bm{k})$, we evaluate the pair coherence length $\xi$, 
which corresponds to the spatial size of the electron-hole pair 
and may be defined by \cite{seki2,ejima,kaneko3} 
\begin{align}
\xi^2
=\frac{\sum_{\bm{k}}|\nabla_{\bm{k}}F(\bm{k})|^2}{\sum_{\bm{k}}|F(\bm{k})|^2}.  
\end{align}
In Fig.~\ref{fig5}, we show the calculated results for the 
spin-singlet excitons ($\xi_0$) and spin-triplet excitons 
($\xi_z$) as a function of $J$.  
We find that, with increasing $J$ ($=J'$), $\xi$ for the spin-singlet 
(triplet) excitons increases (decreases) monotonically.  Thus, the 
size of the spin-singlet exciton becomes larger than the lattice constant 
($\xi_0>1$) for larger $J$ values, indicating the crossover from the 
tightly paired BEC state to the weakly paired BCS state.  
The spin-triplet excitons, on the other hand, are paired more tightly, 
and the size is always smaller than the lattice constant in the 
parameter space examined.  We also find in the inset of Fig.~\ref{fig5} 
that the above tendencies induced by Hund's rule coupling are 
again enhanced by the pair hopping term.

\section{Summary and discussion}

To summarize, we have studied the stability of the excitonic 
density-wave states in the two-band Hubbard model with the 
interorbital Coulomb interaction $U'$, Hund's rule coupling $J$, 
 pair hopping term $J'$, as well as the intraorbital Hubbard 
interaction $U$.  We have rewritten the interorbital interactions 
of the Hamiltonian in terms of the creation and annihilation operators 
of the spin-singlet and spin-triplet excitons and examined the roles 
of these interactions.  We have thereby shown that the $U'$ term 
drives the formation of excitons in both the spin-singlet and the 
spin-triplet channels, and the $J$ term stabilizes (destabilizes) 
the formation of the spin-triplet (spin-singlet) excitons.  
Using the VCA to calculate the grand potential of the system in the 
thermodynamic limit, we have moreover shown that Hund's rule 
coupling always stabilizes the excitonic SDW state and destabilizes 
the excitonic CDW state of which the tendencies are enhanced by 
the pair hopping term.  
A variety of physical quantities has also been calculated, which 
include the single-particle spectral function, density of states, 
anomalous Green's functions, condensation amplitude, and pair 
coherence length.  We have thus characterized the excitonic CDW 
and SDW states in detail.  
\\\indent
Finally, let us discuss the experimental implications of our results 
obtained in this paper.  At first sight, the condensations of the 
spin-singlet excitons possibly observed in $1T$-TiSe$_2$ 
(Refs.~[\onlinecite{cercellier,monney1,monney2,zenker1}]) 
and Ta$_2$NiSe$_5$ (Refs.~[\onlinecite{wakisaka1,wakisaka2,kaneko1,seki1}]) 
seem to contradict the stability of the spin-triplet excitons in 
the presence of Hund's rule coupling.  However, in these 
materials, the valence and conduction bands are formed by the 
orbitals located on different atoms, i.e., the $4p$ orbitals of Se 
ions for the valence bands and the $3d$ orbitals of Ti ions for 
the conduction bands in $1T$-TiSe$_2$,\cite{cercellier} and the 
$3d$ orbitals of Ni ions for the valence bands, and the $5d$ orbitals 
of Ta ions for the conduction bands in Ta$_2$NiSe$_5$,\cite{kaneko1} 
and therefore Hund's rule coupling acting between electrons on 
different orbitals in a single ion does not work to stabilize the 
condensation of the spin-triplet excitons.  We anticipate that in 
these materials the electron-phonon coupling should work to stabilize 
the condensation of the spin-singlet excitons as was discussed in 
Refs.~[\onlinecite{kaneko1,monney2,zenker1}].  
In the excitonic SDW states possibly observed in, e.g., iron pnictide 
superconductors and Co oxide materials, on the other hand,  
Hund's rule coupling rather than the electron-phonon coupling 
should work to stabilize the condensation of the spin-triplet 
excitons as we have shown in this paper.  We may therefore suggest 
that the competition between Hund's rule coupling and electron-phonon 
coupling in the stability of excitonic condensations (or excitonic 
density-wave formations) will be of great interest in future 
studies.  

\begin{acknowledgments}
T.~K.~acknowledges support from the JSPS Research Fellowship for Young Scientists.  
This work was supported, in part, by a Kakenhi Grant No.~26400349 from 
JSPS of Japan.  
\end{acknowledgments}

\begin{figure}[t]
\begin{center}
\includegraphics[width=\columnwidth]{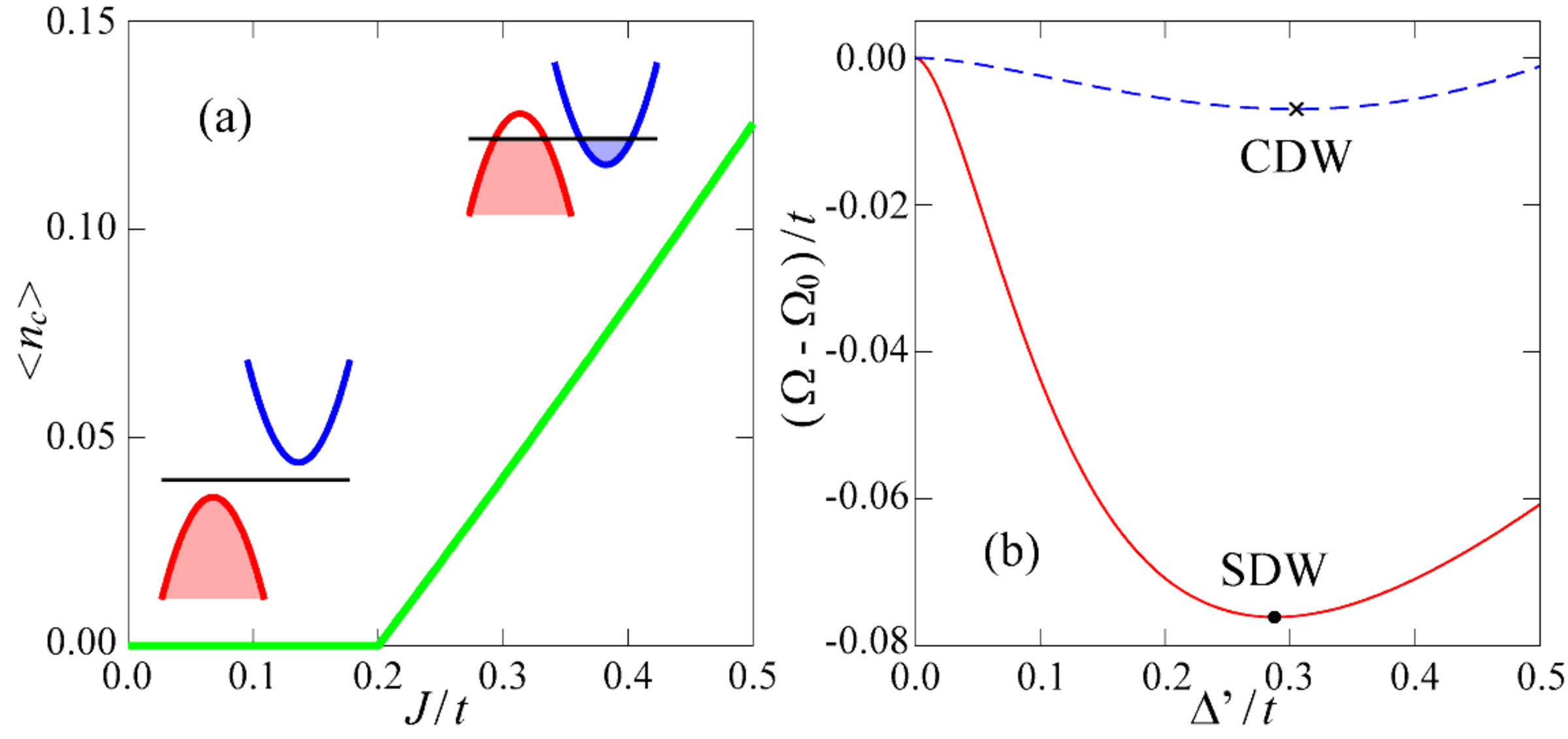}
\caption{(Color online) (a) The number of the conduction-band electrons 
$\langle n_c\rangle$ (or the valence-band holes) as a function of $J/t$ in the normal 
state (or $\Delta'=0$), which is obtained using the atomic-limit relation 
$U'=U-2J$ with $U/t=5$, $D/t=2$, and $J'=0$.  
(b) Calculated grand potentials of the excitonic CDW and SDW states 
as a function of the variational parameter $\Delta'$ ($=\Delta_0',\Delta_z'$), 
which are obtained using the atomic-limit relation $U'=U-2J$ with 
$U/t=5$, $J/t=J'/t=0.5$ and $D/t=2$.  The crosses and circles indicate the 
stationary points of the excitonic CDW and SDW states, respectively.  }
\label{fig6}
\end{center}
\end{figure}

\begin{appendix}
\section{Use of the atomic-limit relation}

In the main text, we have assumed the relation $U'=(U+J)/2$ between 
the interaction parameters.  In this Appendix, we present some results 
obtained in a different choice of the relation, i.e., $U'=U-2J$, which 
is valid in the atomic limit,\cite{griffith,brandow,oles} and show 
that the essential features of our results do not alter.  
\\\indent
In the BCS-like mean-field theory applied to our model Eq.~(\ref{ham}), 
the diagonal terms of the mean-field Hamiltonian are given by 
$\varepsilon_f(\bm{k})=-\varepsilon_c(\bm{k})=-2t\sum_i^d\cos k_i-D+n(U/2-U'+J/2)$ 
with $n = \langle n_{if\sigma} \rangle - \langle n_{ic\sigma} \rangle$, 
and the off-diagonal term gives the spontaneous $c$-$f$ hybridization 
(or excitonic condensation).\cite{kaneko1} 
The Hartree shift $n(U/2-U'+J/2)$ appears in this expression.  
Depending on the values of $U$, $U'$ and $J$, we therefore find, e.g., 
the Mott-insulator state at $U'\ll(U+J)/2$ and the band-insulator state 
at $U'\gg(U+J)/2$,\cite{zocher,kaneko2} which are due simply to 
the effect of the Hartree shift.  
\\\indent
The effects of this Hartree shift can be suppressed completely if we 
assume the relation $U'=(U+J)/2$ as in the main text.  However, if we 
assume the atomic-limit relation $U'=U-2J$, the change in the parameter 
values, e.g., $J$, leads to the change in the overlap of the valence and 
conduction bands and hence to the change in the number of conduction-band 
electrons (and valence-band holes) as shown in Fig.~\ref{fig6}(a).  
This gives an additional complexity to our calculations because in this 
paper we just want to focus on the relative stability of the excitonic 
CDW and SDW states in the presence of Hund's rule coupling and the pair 
hopping term.  
\\\indent
Our assumption of the use of the relation $U'=(U+J)/2$ may be justified 
if the essential features obtained in the main text do not differ from 
the results obtained using the atomic-limit relation $U'=U-2J$.  
In Fig.~\ref{fig6}(b), we show the grand potentials as a function of 
the variational parameter $\Delta'$ calculated using the atomic-limit 
relation $U'=U-2J$ where we actually find that the results are nearly 
the same as the results shown in Fig.~\ref{fig1}(a) in the main text.  
Therefore, we may safely state that the essential features obtained in 
the main text do not alter in the different choice of the parameter set.  
\end{appendix}

\end{document}